\documentstyle{mn}

\input epsf

\begin{document}

\def\Ha{H$\alpha\ $}
\def\Hb{H$\beta\ $}
\def\Lya{Ly$\alpha\ $}
\def\Lyb{Ly$\beta\ $}
\def\Lyg{Ly$\gamma\ $}
\def\Lyd{Ly$\delta\ $}
\def\Lye{Ly$\epsilon\ $}
\def\LCDM{$\Lambda$CDM\ }
\def\HI{\hbox{H~$\rm \scriptstyle I\ $}}
\def\HII{\hbox{H~$\rm \scriptstyle II\ $}}
\def\DI{\hbox{D~$\rm \scriptstyle I\ $}}
\def\HeI{\hbox{He~$\rm \scriptstyle I\ $}}
\def\HeII{\hbox{He~$\rm \scriptstyle II\ $}}
\def\HeIII{\hbox{He~$\rm \scriptstyle III\ $}}
\def\CII{\hbox{C~$\rm \scriptstyle II\ $}}
\def\CIII{\hbox{C~$\rm \scriptstyle III\ $}}
\def\CIV{\hbox{C~$\rm \scriptstyle IV\ $}}
\def\NV{\hbox{N~$\rm \scriptstyle V\ $}}
\def\OIII{\hbox{O~$\rm \scriptstyle III\ $}}
\def\OIV{\hbox{O~$\rm \scriptstyle IV\ $}}
\def\OVI{\hbox{O~$\rm \scriptstyle VI\ $}}
\def\SiIV{\hbox{Si~$\rm \scriptstyle IV\ $}}
\def\NHI{N_{\rm HI}}
\def\NHeII{N_{\rm HeII}}
\def\cm2{\,{\rm cm$^{-2}$}\,}
\def\kms{\,{\rm km\,s$^{-1}$}\,}
\def\skm{\,({\rm km\,s$^{-1}$})$^{-1}$\,}
\def\kmsmpc{\,{\rm km\,s$^{-1}$\,Mpc$^{-1}$}\,}
\def\hmpc{\,h^{-1}{\rm \,Mpc}\,}
\def\mpch{\,h{\rm \,Mpc}^{-1}\,}
\def\hkpc{\,h^{-1}{\rm \,kpc}\,}
\def\ev{\,{\rm eV\ }}
\def\kel{\,{\rm K\ }}
\def\intunits{\,{\rm ergs\,s^{-1}\,cm^{-2}\,Hz^{-1}\,sr^{-1}}}
\def\ltsima{$\; \buildrel < \over \sim \;$}
\def\lsim{\lower.5ex\hbox{\ltsima}}
\def\gtsima{$\; \buildrel > \over \sim \;$}
\def\gsim{\lower.5ex\hbox{\gtsima}}
\def\etal{{ et~al.~}}
\def\aj{AJ}
\def\ana{A\&A}
\def\apj{ApJ}
\def\apjs{ApJS}
\def\mn{MNRAS}

\journal{Preprint-00}

\title{The possible detection of high redshift Type II QSOs in deep fields}

\author[A. Meiksin]{Avery Meiksin \\
SUPA\thanks{Scottish Universities Physics Alliance},
Institute for Astronomy, University of Edinburgh,
Blackford Hill, Edinburgh\ EH9\ 3HJ, UK}

\pubyear{2005}

\maketitle

\begin{abstract}
The colours of high redshift Type II QSOs are synthesized from
observations of moderate redshift systems. It is shown that Type II
QSOs are comparable to starbursts at matching the colours of
$z_{850}$-dropouts and $i_{775}$-drops in the Hubble Ultra Deep Field,
and more naturally account for the bluest objects detected. Type II
QSOs may also account for some of the $i_{775}$-drops detected in the
GOODS fields. It is shown that by combining imaging data from the {\it
Hubble Space Telescope} and the {\it James Webb Space Telescope}, it
will be possible to clearly separate Type II QSOs from Type I QSOs and
starbursts based on their colours. Similarly, it is shown that the
UKIDSS $ZYJ$ filters may be used to discriminate high redshift Type II
QSOs from other objects. If Type II QSOs are prevalent at high
redshifts, then AGN may be major contributors to the re-ionization of
the Intergalactic Medium.
\end{abstract}


\section{Introduction} \label{sec:introduction}

The band-dropout method has proven an impressive means for discovering
high redshift objects. Pioneered by Guhathakurta \etal (1990), Bithell
(1991), and Steidel \& Hamilton (1992, 1993), the method exploits the
downward step in flux shortward of the Lyman limit for discovering
starburst galaxies. Absorption by intervening hydrogen in the
Intergalactic Medium (IGM) will similarly produce extremely red
optical colours for objects at $z\gsim3$, a characteristic signature
of a high redshift system (Bithell 1991; Madau 1995). Following
Steidel \& Hamilton's discovery of a population of Lyman break
galaxies at $z\approx3$, the method has evolved into a standard tool
for identifying even higher redshift objects in recent very deep
surveys like the Hubble Deep Field and the Ultra-Deep Field. While
follow-up spectroscopy has demonstrated that most of the objects are
galaxies, a few dozen reveal the emission line signatures of Active
Galactic Nuclei (AGN) (Steidel \etal 2002), about half of which have
the narrow lines characteristic of Type~II systems. These AGN have
made possible the first measurement of the faint end of the luminosity
function of Quasi-Stellar Objects (QSOs) at $z\approx3$ (Hunt \etal
2004), critical for assessing the poorly constrained contribution of
QSOs to the UV ionizing background at this epoch (Meiksin 2005a).

Since then, band-dropout objects have been discovered in a host of
surveys, most recently pushing to $z\approx6$ (Stanway, Bunker \&
McMahon 2003; Bouwens \etal 2004; Bunker \etal 2004; Yan \& Windhorst
2004; Giavalisco \etal 2004). The objects have generally been modelled
as young star-forming galaxies and used to infer the cosmic star
formation rate of the universe and its evolution. They have also been
used to assess the contribution of galaxies to the budget of ionizing
photons required to re-ionize the universe, concluding their numbers
are either too few (Bunker \etal 2004) or easily adequate (Stiavelli
\etal 2004), depending on model assumptions.

In this paper, it is suggested that a portion of the band-dropout
objects may be Type II QSOs.\footnote{There is not universal consensus
on the definition of a Type II QSO. In this paper, a definition
similar to that of Zakamska \etal (2003) is assumed, that the
restframe FWHM of hydrogen lines be less than 2000\kms. An alternative
x-ray motivated definition is based on x-ray spectral evidence for a
large obscuring \HI column density local to the active nucleus, for
example, Gandhi \etal (2004).} The prospects of detecting Type II QSOs
in deep surveys has not previously been explored. These objects have
been of considerable recent interest because they are predicted in
unification models of AGN (Antonucci 1993). Discovering examples at
higher redshifts would help to further elucidate their properties and
their connection to Type I QSOs. As sources of energetic photons, they
are also candidate sources for high energy background
radiations. Obscured QSOs have been postulated as sources of the hard
x-ray background (Madau, Ghisellini \& Fabian 1994), and Type II QSOs
are natural candidates for such sources. While Type I QSOs may
dominate the UV ionizing metagalactic background at $z\lsim3.5$, their
numbers appear inadequate at higher redshifts (eg, Meiksin
2005a). Although galaxies are possible sources, the abundance of high
redshift Type II QSOs is far too uncertain to rule out an ionizing
background dominated by AGN sources. If their numbers are sufficiently
great at high redshifts, Type II QSOs may also have contributed
substantially to the re-ionization of the IGM. Indeed, if the IGM were
re-ionized at $z=6-8$, consistency with measurements of the optical
depth of the IGM at $z<6$ favour hard ionizing sources like Pop~III
stars and AGN (Meiksin 2005a).

The Sloan Digital Sky Survey (SDSS) has discovered nearly 300 Type II
QSOs over the redshift range $0.3<z<0.83$, with the rest wavelengths
of the measured spectra restricted to being longward of about
2100~\AA\ (Zakamska \etal 2003). These elusive narrow-lined objects
have proven even more difficult to discover at higher redshifts, and
their UV spectral properties are consequently still poorly
understood. To date, only four Type II QSOs are known to exist at
$z>1.5$ with optical spectra showing narrow emission lines, two
discovered through observations with the {\it Chandra X-Ray
Observatory} (Norman \etal 2002; Mainieri \etal 2005), one discovered
as an optical band-dropout (Stern \etal 2002), and one discovered
through optical spectroscopy (Jarvis \etal 2005). A fifth candidate
was found by Dawson et al (2001), but with incomplete emission line
results reported. The discovery of high redshift counterparts would
add substantially to our knowledge of the nature and origin of these
systems.

\section{Source models}

To more clearly delineate the differences between the predicted
colours of Type II QSOs and other possible sources, the colours of
starburst galaxies and Type I QSOs are also considered. Stanway,
McMahon \& Bunker (2005) previously considered starbursts as the
origin of their $i_{775}$-drops found in the Hubble Ultra-Deep Field
(UDF), and found they were unable to match the colours of the bluest
sources in $(J_{110}-H_{160})_{\rm AB}$, including a population of
blue sources too dim to measure in $J_{110}$ but for which colours are
inferred from image stacking. Their results are confirmed and slightly
extended here. Bouwens \etal discovered $z_{850}$-dropouts in the UDF
with similarly blue $(J_{110}-H_{160})_{\rm AB}$ colours. The
$i_{775}$-drops detected in the UDF by Bunker \etal (2004) and Yan \&
Windhorst (2004) are also considered, as are the $i_{775}$-drops
reported by Stanway, Bunker \& McMahon (2003) and Eyles \etal (2005)
as part of the Great Observatories Origins Deep Survey (GOODS).

The starbursts models used here were generated using the STARBURST99
package of Leitherer \etal (1999). Continuous star formation is
assumed with a Salpeter Initial Mass Function. Nebular emission lines
are included. Solar metallicity is assumed, although the colour
results below are not very sensitive to the metallicity. For instance,
using a metallicity of 0.05 solar reduces $(z_{850}-J_{110})_{\rm AB}$
for a 30~Myr starburst at $z\approx6$ by less than 0.01~mag below the
value for solar metallicity and $(J_{110}-H_{160})_{\rm AB}$ by less
than 0.1~mag. Colours are presented in the figures below for ages of
30~Myr and 600~Myr, which is just under the age of the universe at
$z\approx8.5$ for a cosmology with $\Omega_M=0.3$, $\Omega_v=0.7$, and
$h=0.7$.

The Type I QSO spectrum adopted is the median spectrum constructed by
Vanden Berk \etal (2001) from the SDSS QSO survey. It is a composite
of over 2200 spectra homogeneously selected from the SDSS QSO survey
covering the redshift range $0.044\leq z\leq 4.789$. The spectrum
covers the rest wavelength range $800-8555$~\AA. While the colours of
Type I QSOs will show a fairly wide spread, as investigated by Chiu
\etal (2005), the median composite is a good indicator of the locus of
a Type I QSO in colour space.

Since no flux-calibrated spectra of Type II QSOs exist that clearly
reveal the continuum over an extended wavelength range, a Type II
prototype spectrum is constructed from measured broad band magnitudes
and emission lines. In addition to having narrower emission lines,
there is an indication in the literature that some Type II QSOs differ
from Type I QSOs in another fundamental aspect:\ several of the
emission lines appear often to have high equivalent widths comparable
to the filter bandwidths. This will give them unusual colours,
especially at high redshifts, since the emission lines will dominate
the continuum light in a given passband. It is the observational
consequences of this particular feature of some Type II QSOs that is
emphasized in this paper, although low equivalent width Type II QSOs
are also considered.

Estimating the contribution of the emission lines to the Type II QSO
colours, however, has several uncertainties. The spectra and
photometry are measured through different size apertures. This
requires an aperture correction for which an arbitrary assumption must
be made regarding any possible spatial variation of the light in the
emitting regions. Norman \etal\ show substantial corrections to the
measured magnitudes of a Type II QSO after removing the emission
lines. Unfortunately they do not indicate the size of the aperture
correction they must have used. The weaknesses of the reported flux
values of the emission lines compared with the broadband-integrated
fluxes show, however, that an aperture correction of $\sim20$ must
have been applied, which seems rather high. The Mainieri \etal QSO has
low equivalent width emission lines, so that its spectrum will differ
little from that of an AGN-hosting ULIRG like NGC~6240 when boosted to
higher redshifts, which well matches the colours measured by Mainieri
\etal The Jarvis \etal QSO is at too low a redshift to detect
Ly$\alpha$.

\begin{figure}
\begin{center}
\leavevmode \epsfxsize=3.3in \epsfbox{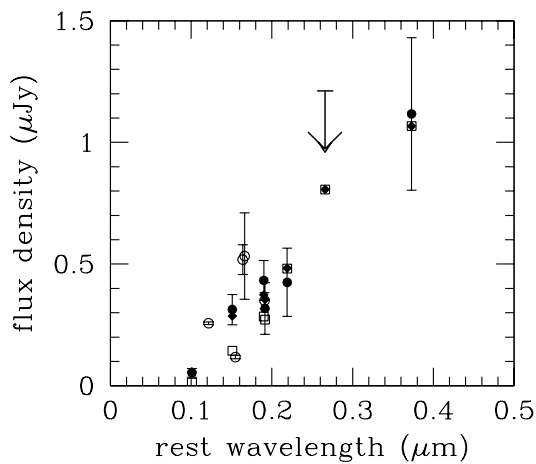}
\end{center}
\caption{The measured and synthesized flux densities for the Type II
QSO CXO~52. The measured flux densities in each band from Stern \etal
(2002) are shown by the filled points with error bars, and one upper
limit. The wavelengths are in the rest frame of the QSO. The inferred
un-attenuated continuum of the QSO is shown by the open squares. The
predicted flux densities, including the contribution from the emission
lines and the effect of intergalactic attenuation, are shown by the
filled diamonds. The open points with error bars show the
aperture-corrected continuum levels inferred from the measured fluxes
and equivalent widths of the optical emission lines reported by Stern
\etal
}
\label{fig:mags}
\end{figure}

\begin{table}
\begin{center}
\begin{tabular}{|l|c|c|c|} \hline
\hline 
& Flux & $W_\lambda$ (obs) & $W^{a.c.}_\lambda$ (obs) \\
Line & ($10^{-17}\,{\rm ergs\,cm^{-2}\,s^{-1}}$)  & (\AA) & (\AA) \\
\hline
\OVI~$\lambda1035$ & 1.5: & 120: & 345 \\
\Lya & $18.9\pm0.4$ & $2100\pm40$ & 1948 \\
\NV~$\lambda1240$ & 0.6: & 50: & 56 \\
\SiIV/\OIV]~$\lambda1403$ & 0.4: & 30: & 20 \\
\CIV~$\lambda1549$ & $3.5\pm0.2$ & $350\pm20$ & 676 \\
\HeII~$\lambda1640$ & $1.7\pm0.2$ & $170\pm20$ & 361 \\
\OIII~$\lambda1663$ & $0.9\pm0.3$ & $90\pm30$ & 196 \\
\CIII]~$\lambda1909$ & $2.1\pm0.2$ & $420\pm40$ & 574 \\
\hline
\end{tabular}
\end{center}
\caption{Emission line measurments of CXO~52 from Stern \etal (2002).
The first equivalent width column is as measured through a 1.5\arcsec
x 1.5\arcsec~ aperture; the second column is aperture-corrected to a
3\arcsec~ diameter aperture (see text). Parameters with colons
indicate uncertain measurements. The equivalent width values are in
the observer's frame.
}
\label{tab:emlines}
\end{table}

Only the Stern \etal QSO, CXO~52 ($z=3.288$) provides all the
requirements for constructing a Type II QSO spectrum in the UV with
high equivalent width emission lines. The emission line properties
they fit are shown in Table~\ref{tab:emlines}. It should be noted that
the published Table~2 in Stern \etal contains a misprint in the column
heading for the equivalent widths, mistakenly indicating they are in
the rest frame when in fact they are in the observer's frame (D. Stern
personal communication). The emission lines were measured from a
spectrum taken through a 1.5\arcsec x 1.5\arcsec~ aperture. The
magnitudes reported are for a 3\arcsec~ diameter aperture. Without
detailed surface brightness profiles of the continuum and emission
line emitting regions, it is not possible to make an assumption-free
comparison of the overall contribution of the emission lines to the
broadband fluxes. In particular, the optical broadband flux densities
in Table 1 of Stern \etal indicate flux levels of about $0.3\,\mu{\rm
Jy}$, while the optical spectrum in their Figure 3 shows a continuum
level of about $0.1\,\mu{\rm Jy}$. Except possibly for Ly$\alpha$, the
quoted equivalent widths in their Table 2 are too small for the
emission lines to dominate the broadband optical magnitudes, so that
the difference in the broadband and spectral continuum flux levels
indicates that the continuum extends beyond the 1.5\arcsec x
1.5\arcsec~ slit.  It is consequently necessary to make some
assumptions regarding the spatial distribution of the emission line
and continuum emitting regions to infer the contribution of the
emission lines to the broadband colours.

The \Lya emitting region in Figure 5 of Stern \etal clearly extends
beyond the 1.5\arcsec x 1.5\arcsec~ aperture. Although the
emission-line region may peak within the 1.5\arcsec x 1.5\arcsec~
aperture, its surface brightness distribution is unknown. If shallow,
it is still possible that the integrated flux of the emission-line
region continues to increase substantially beyond the slit. The
assumption adopted here is that the line-emitting region fills the
3\arcsec~ aperture used for the broadband magnitudes, and an aperture
correction is applied assuming uniform surface brightness.

The aperture-corrected emission line fluxes are then removed from the
broadband magnitude measurements to infer the levels of the underlying
continuum. The spectrum is then constructed by fitting power-laws
between the values of the continuum at the effective wavelengths of
each filter, adjusting the values to match the measured magnitudes
after re-introducing the emission lines. Because some of the magnitude
measurements correspond to rest wavelengths shortward of Ly$\alpha$,
the spectrum is corrected for IGM absorption using the values from
Meiksin (2005b). The shortest wavelengths are synthesized by extending
the UV portion of the spectrum to the x-ray using the measured {\it
Chandra} flux in the $0.5-2$~keV band (Stern \etal). The intrinsic and
predicted magnitudes, including IGM attenuation, are compared with
those measured in Figure~\ref{fig:mags}. Also shown are the continuum
levels inferred from the measured fluxes and equivalent widths of the
optical emission lines reported by Stern \etal The continuum levels
are shown at the line centres and corrected to a 3\arcsec\ aperture.
Agreement with the continuum levels inferred from the emission-line
subtracted broadband fluxes is good, although the emission
line-inferred continuum levels are somewhat high near restframe
$\lambda\approx0.16\mu$m.

In terms of the relative contribution of the emission lines to the
continuum, the above procedure is equivalent to assuming a constant
surface brightness for the continuum, aperture-corrected downwards to
the 1.5\arcsec x 1.5\arcsec~ spectroscopic slit, although this would
then require the corresponding correction be applied to the quoted
total broadband magnitudes. It should be clear that the total
magnitudes being presented here are based on apertures that are filled
by the emission-line emitting regions. Should the continuum extend
well beyond the emission line regions, then magnitudes measured
through apertures larger than the emission-line regions will be more
weighted toward the continuum. As the apertures used for broadband
optical magnitude measurements with {\it HST} typically have diameters
of 0.5\arcsec - 1.0\arcsec, this is not generally expected to be a
major complication, although it will obviously depend on the
properties of each source.

The \Ha and \OIII$\lambda\lambda4959, 5007$ lines detected by Stern
\etal lie within the $K_s$ band for the QSO. These lines, after
aperture-correcting, are, within the errors, able to account fully for
the measured $K_s$ magnitude. It is consequently not possible to make
a firm continuum estimate for the QSO at these wavelengths, nor a
reliable estimate for the aperture-corrected equivalent widths of
these lines. Although $K_s$ predictions are made below at higher
redshifts, these predictions are quite uncertain for $z\lsim4$, being
based almost entirely on the aperture-corrected fluxes of these three
lines. Since the longest wavelength near-infra-red continuum point
that is firmly established is based on observations through the
$F160W$ filter, the $K_s$ predictions are in fact fairly uncertain up
to $z\lsim5.3$.

The equivalent widths inferred for the emission lines in a 3\arcsec~
aperture are provided in Table~\ref{tab:emlines}. It should be noted
that while the equivalent widths are similar to those reported by
Stern et al., a few are larger. This may be a consequence of the
aperture correction applied to the emission lines. Under the
alternative view of aperture correcting the broadband continuum
magnitudes from the 3\arcsec~ diameter aperture to the 1.5\arcsec x
1.5\arcsec~ slit, the equivalent widths would be expected to be the
same. It is possible that the equivalent width errors quoted in Table
2 of Stern \etal are underestimates, as they appear only to account
for the errors in the measured emission line fluxes, not for the
errors in the fit continuum levels. The spectrum shown in their Figure
3 suggests the continuum is comparable to the noise level, so that the
continuum error may be large. In any case, the purpose here is not to
provide a definitive spectrum, but a plausible one consistent with the
data. The effect of varying the strengths of the emission lines on the
predicted colours will be considered below to allow for the
uncertainty in their equivalent widths, as well as possible source
metallicity variations.

As a contrast to CXO~52, a spectrum for the Type II QSO CDFS-263 at
$z=3.660$ with low equivalent width emission lines discovered by
Mainieri \etal is also synthesized, using exactly the same procedure
as for CXO~52. This spectrum will be referred to as the ``low
equivalent width'' Type II QSO; otherwise colours based on CXO~52 are
generally meant when referring to Type II QSOs.

\section{High redshift colour predictions}

\subsection{Comparison with Ultra-Deep Field $z_{850}$-dropouts}

\begin{figure}
\begin{center}
\leavevmode \epsfxsize=4in \epsfbox{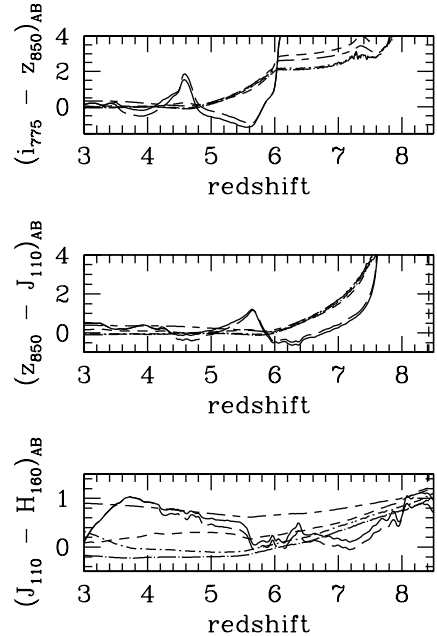}
\end{center}
\caption{The redshift evolution of $(i_{775} - z_{850})_{\rm AB}$,
$(z_{850}-J_{110})_{\rm AB}$ and $(J_{110}-H_{160})_{\rm AB}$ colours
predicted for starburst galaxies of age 30~Myr (dotted long-dashed
lines) and 600~Myr (dotted short-dashed lines), Type I QSO
(short-dashed lines), and high equivalent width Type II QSO (solid
lines). The long-dashed line shows the track for the Type II QSO with
\CIV enhanced by a factor of 2. The blueing in
$(i_{775}-z_{850})_{AB}$ for the Type II QSO at $z>4.6$ is due to the
entering of the \Lya emission line into the $i_{775}$ passband, which
it then leaves by $z\gsim6$.  Similarly, the blueing in
$(z_{850}-J_{110})_{AB}$ for the Type II QSO at $z>5.7$ is due to the
entering of the \Lya emission line into the $z_{850}$ passband, which
it then leaves by $z=7.6$. The short-dashed long-dashed lines show the
results for a low equivalent width Type II QSO.
}
\label{fig:colourev}
\end{figure}

\begin{figure}
\begin{center}
\leavevmode \epsfxsize=3.3in \epsfbox{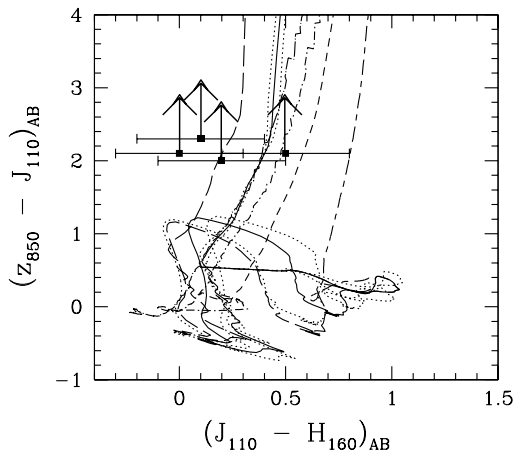}
\end{center}
\caption{$(z_{850}-J_{110})_{\rm AB}$ vs $(J_{110}-H_{160})_{\rm AB}$
colour-colour diagram showing the position of the $z_{850}$-dropouts
of Bouwens \etal and the predicted tracks at $z>3$ for starburst
galaxies of age 30~Myr (dotted long-dashed line) and 600~Myr (dotted
short-dashed line), Type I QSO (short-dashed line), and Type II QSO
with high equivalent width emission lines (solid line). The dotted
lines show the tracks for the Type II QSO with the metal lines boosted
or reduced by 50\%. The long-dashed line shows the track for the Type
II QSO with \CIV enhanced by a factor of 2. The short-dashed
long-dashed line shows the predictions for a low equivalent width Type
II QSO.
}
\label{fig:zJ-JH}
\end{figure}

The evolution of the predicted $(i_{775}-z_{850})_{\rm AB}$,
$(z_{850}-J_{110})_{\rm AB}$ and $(J_{110}-H_{160})_{\rm AB}$ colours
of the candidate sources are shown in Figure~\ref{fig:colourev}. The
entrance of the \Lya emission line of the high equivalent width Type
II QSO into the $z_{850}$ passband (middle panel) is visible as a
sharp blueing for $z\gsim5.7$. The object becomes a $z_{850}$-dropout
quite abruptly at $z>7.6$, at which point it is the bluest of the
candidate sources in $(J_{110}-H_{160})_{\rm AB}$.

Bouwens \etal (2004) detect four $z_{850}$-dropouts in the Hubble
Ultra-Deep Field, interpreting them as starbursts in the redshift
range $7<z<8$, to which the $z_{850}$-dropout method is sensitive
through the detection of the \Lya break. These objects are unusually
blue in $(J_{110}-H_{160})_{\rm AB}$, bluer than predicted for
starbursts starbursts, even when the light is dominated by Pop~III
objects. To model the colours, the ACS $F850LP$ filter transmission is
combined with the total quantum efficiency of the detector to compute
$z_{850}$. Similarly, the NICMOS $F110W$ and $F160W$ filters are used
to compute $J_{110}$ and $H_{160}$. The AB scale is assumed
throughout. The effects of intergalactic attenuation are included
(Meiksin 2005b). The $(z_{850}-J_{110})_{\rm AB}$ vs
$(J_{110}-H_{160})_{\rm AB}$ colours are shown for the two starburst
models and the Type I and Type II QSO spectra in
Figure~\ref{fig:zJ-JH}. The Type I QSO, the low equivalent width Type
II QSO and the 600~Myr starburst galaxy colours poorly match the
$z_{850}$-dropout colours. The large equivalent width Type II QSO and
30~Myr starburst are equally good matches.

Because of the uncertain equivalent widths of the metal emission lines
in CXO~52, the effect of varying them is explored. This may also be
considered a change in the overall metallicity, although it ignores
any effect changing the metallicity may have on the thermodynamics of
the emitting gas and hence the emission line strengths. For
$(z_{850}-J_{110})_{\rm AB}<1.8$, boosting the metallicity by 50\%
renders the Type II QSO bluer in $(J_{110}-H_{160})_{\rm AB}$, while
it is reddened for $(z_{850}-J_{110})_{\rm AB}>1.8$. Decreasing the
metallicity by 50\% has the opposite effect. The effect on the
$z_{850}$-dropouts is small, as shown in Figure~\ref{fig:zJ-JH}.

The emission line flux ratios for CXO~52 are consistent with the
photoionization models presented in Table~6 of Kwan \& Krolik (1981)
for clouds exposed to an AGN spectrum. These models predict that most
of the carbon is in the form of \CII, and that increasing the density
of the clouds tends to increase the emission ratio of
\CIV$\lambda1549$ to \CIII]$\lambda1909$. A model with the \CIV
equivalent width doubled, holding all else fixed, is shown in
Figure~\ref{fig:zJ-JH}. The colour track now runs through the
$z_{850}$-dropout points. Although it is not possible to rule out
statistically the possibility that all four objects are starbursts,
this does show that Type II QSOs are plausible candidates. In
particular, the three bluest objects, taken collectively with an
average $(J_{110}-H_{160})_{\rm AB}\approx0.1\pm0.2$, are consistent
with a 30~Myr starburst and 600~Myr starburst only at the 1.5$\sigma$
and 2$\sigma$ levels, respectively, favouring the interpretation that
one or more of these three is a Type II QSO with large equivalent
widths.

\begin{figure}
\begin{center}
\leavevmode \epsfxsize=3.3in \epsfbox{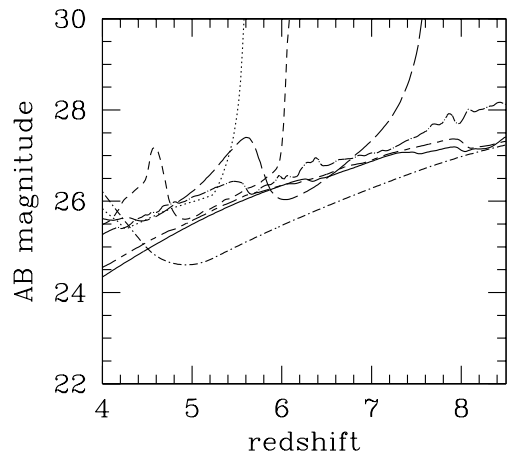}
\end{center}
\caption{$i_{775}$ (short-dashed line), $z_{850}$ (long-dashed line),
$J_{110}$ (dotted long-dashed line), $H_{160}$ (solid line) and $K_s$
(dotted short-dashed line) AB magnitudes predicted for the Type II QSO
CXO~52. Also shown are the predicted magnitudes that would be measured
through the $F070W$ (dotted line) and $F150W$ (short-dashed
long-dashed line) filters by {\it JWST}. The $K_s$ magnitude
predictions are uncertain for $z\lsim5$ (see text).
}
\label{fig:magev}
\end{figure}

The predicted evolution of the magnitudes in various passbands of
CXO~52 is shown in Figure~\ref{fig:magev}. The predicted $H_{160}$
magnitude at $z>6$ is fully consistent with the measured values
reported in Bouwens \etal of $H_{160}=26.0-27.1$.

\subsection{Comparison with deep field $i_{775}$-drops}

\begin{figure}
\begin{center}
\leavevmode \epsfxsize=3.3in \epsfbox{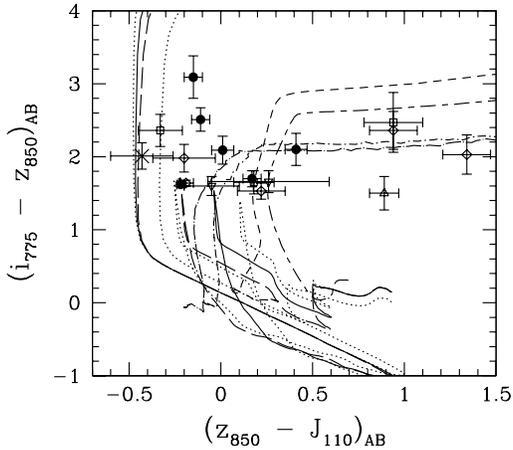}
\end{center}
\caption{$(i_{775}-z_{850})_{\rm AB}$ vs $(z_{850}-J_{110})_{\rm AB}$
colour-colour diagram predicted at $z>3$ for starburst galaxies of age
30~Myr (dotted long-dashed line) and 600~Myr (dotted short-dashed
line), Type I QSO (short-dashed line), and Type II QSO with high
equivalent width emission lines (solid line). The dotted lines show
the tracks for the Type II QSO with the metal lines boosted or reduced
by 50\%. The long-dashed line shows the track for the Type II QSO with
\CIV enhanced by a factor of 2. The short-dashed long-dashed line
shows the predictions for a low equivalent width Type II QSO.  The
locus for the high equivalent width Type II QSO with metal lines cut
by 50\% lies redward of the solid line in $(z_{850}-J_{110})_{\rm AB}$
for $(i_{775}-z_{850})_{\rm AB}<1.5$, while the bluest dotted line in
$(z_{850}-J_{110})_{\rm AB}$ corresponds to the Type II QSO with the
metals boosted by 50\%. The open diamonds and squares and the `X' are
from measurements of sources in the UDF by Stanway, McMahon \& Bunker
(2005). The diamonds are for individual sources, while the squares
indicate the composite colours of groups of sources. The `X' shows the
colours of stacked source images undetected in $J_{110}$.  The solid
points are from UDF sources identified by Yan \& Windhorst
(2004). Also shown are the colours of GOODS sources measured by
Stanway, Bunker \& McMahon (2003) (open triangles) and Eyles \etal
(2005) (inverted triangles).
}
\label{fig:iz-zJ_idrop}
\end{figure}

\begin{figure}
\begin{center}
\leavevmode \epsfxsize=3.3in \epsfbox{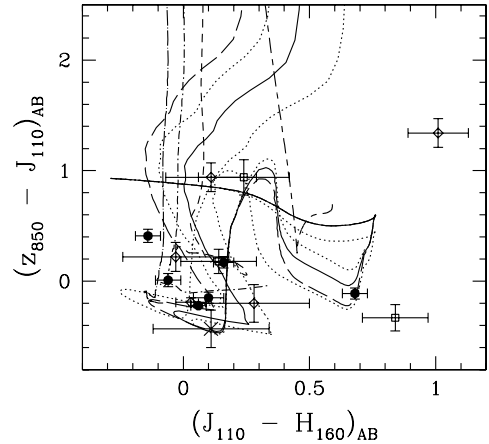}
\end{center}
\caption{$(z_{850}-J_{110})_{\rm AB}$ vs $(J_{110}-H_{160})_{\rm AB}$
colour-colour diagram predicted at $z>3$ for starburst galaxies of age
30~Myr (dotted long-dashed line) and 600~Myr (dotted short-dashed
line), Type I QSO (short-dashed line), and Type II QSO with high
equivalent width emission lines (solid line). The dotted lines show
the tracks for the Type II QSO with the metal lines boosted or reduced
by 50\%. The long-dashed line shows the track for the Type II QSO with
\CIV enhanced by a factor of 2. The short-dashed long-dashed line
shows the predictions for a low equivalent width Type II QSO.  The
locus for the high equivalent width Type II QSO with metal lines
boosted by 50\% lies blueward of the solid line in
$(J_{110}-H_{160})_{\rm AB}$ for $(z_{850}-J_{110})_{\rm AB}>1.1$.
The open diamonds and squares and the `X' are from measurements of
sources in the UDF by Stanway, McMahon \& Bunker (2005). The diamonds
are for individual sources, while the squares indicate the composite
colours of groups of sources. The `X' shows the colours of stacked
source images undetected in $J_{110}$.  The solid points are from UDF
sources identified by Yan \& Windhorst (2004).
}
\label{fig:zJ-JH_idrop}
\end{figure}

A search for lower redshift starbursts in the UDF was conducted by
Bunker \etal (2004) through the identification of $i_{775}$-drops,
defined as objects with $(i_{775}-z_{850})_{\rm AB}>1.3$. The method
is sensitive to galaxies in the redshift range $5.8<z<7$ through
detection of the \Lya break. Follow-up {\it HST} NICMOS observations
were reported by Stanway, McMahon \& Bunker (2005). Objects with
complete $i_{775}$, $z_{850}$, $J_{110}$ and $H_{160}$ photometry are
plotted in Figures~\ref{fig:iz-zJ_idrop} and
\ref{fig:zJ-JH_idrop}. Also shown are data from a similar study by Yan
\& Windhorst (2004). The colour tracks of the various sources
considered in this paper are shown for comparison.

The evolution of the predicted $(i_{775}-z_{850})_{\rm AB}$ colour of
the candidate sources is shown in the upper panel of
Figure~\ref{fig:colourev}. (Not shown are the colours of standard
Hubble type galaxies. These are provided by Stanway \etal 2005.)  The
entrance of the \Lya emission line of the high equivalent width Type
II QSO into the $i_{775}$ passband is visible as a sharp blueing at
$z\gsim4.6$.  This, combined with the rapidly declining flux shortward
of restframe $\lambda<1400$\AA, results in a narrow spur in
$i_{775}-z_{850}$ of redshift width $\Delta z\approx0.1$, as shown in
Figures~\ref{fig:colourev} and \ref{fig:iz-zJ_idrop}. Although it
would seem unlikely to find a Type II QSO in the narrow redshift range
near $z=4.6$ required for it to be an $i_{775}$-drop, it should be
borne in mind that the colours are based on a single object.  The
continuum shape and \Lya emission line strength of a different Type II
QSO would in general result in a differently shaped and positioned
spur, so that the actual redshift width of Type II QSOs that are
$i_{775}$-drops may be larger. All the objects, however, will become
$i_{775}$-dropouts quite abruptly at $z>6$ when the \Lya emission line
leaves the $i_{775}$ band, as shown in Figures~\ref{fig:colourev} and
\ref{fig:magev}.

A few of the detected objects in Figures~\ref{fig:iz-zJ_idrop} and
\ref{fig:zJ-JH_idrop} lie blueward in $(z_{850}-J_{110})_{\rm AB}$ of
the expected tracks for starburst galaxies, although reddening of the
starburst tracks, which will be stronger in $(i_{775}-z_{850})_{\rm
AB}$ than $(z_{850}-J_{110})_{\rm AB}$, may accout for part of the
discrepancy. Some of these lie near the colour spurs predicted for
Type II QSOs at $z\approx4.6$. The corresponding $(J_{110}-H_{160})$
colours, however, are generally inconsistent with a Type II QSO
interpretation.

None the less, several of the source identities have ambiguous
interpretations based on colours alone. The three bluest points in
$(z_{850}-J_{110})_{\rm AB}$ from Stanway \etal (2005) shown in
Figures~\ref{fig:iz-zJ_idrop} and \ref{fig:zJ-JH_idrop} are objects
20104, 25941 and Group 2. None of these readily match the colours of
galaxies (although object 25941 is not badly discrepant given its
large error bars). The colours of object 20104 are consistent with a
Type II QSO with large equivalent width emission lines at $z=6.0$,
although the narrow redshift range for $1.3<(i_{775}-z_{850})_{\rm
AB}<4$ of $\Delta z\approx0.1$ (Figure~\ref{fig:colourev}) suggests
this is unlikely on space density considerations. The same is true of
source 25941, with colours marginally matching the colours for a Type
II large equivalent width QSO in the narrow $(i_{775}-z_{850})_{\rm
AB}$ spur at $z=4.6$. It also marginally matches the colours for a
Type I QSO at $z\approx6$. Additionally, source 23516 matches the
colours of a Type I QSO at $z\approx5.8$. The colours of Group 2 have
no easy QSO interpretation, nor galactic for that matter; the
composite colours appear to be a mix. In addition, the colours of the
objects in Group 3 may indicate the group is dominated by a Type I
QSO. None of the objects identified by Yan \& Windhorst are consistent
with the colours expected for QSOs.

An earlier search for $i_{775}$-drops was made by Stanway, Bunker \&
McMahon (2003) using GOODS data, with additional $i_{775}$-drops
listed in Eyles \etal (2005) accompanied by {\it Spitzer} data. The
reported colours of these sources are plotted in
Figure~\ref{fig:iz-zJ_idrop}. The three points with measured
$(z_{850}-J)_{\rm AB}$ colours all lie near the tips of the spurs for
Type II QSOs, and so are equally good matches to Type II QSO colours
as to starburst or Type I QSO colours. From Figure~\ref{fig:magev},
$(J-K_s)_{\rm AB}\approx1$ is predicted, in good agreement with the
measured values for two of the objects (Eyles \etal). In these cases,
however, follow-up spectroscopy revealed an emission line in each case
corresponding to \Lya at $z\approx5.8$ (Stanway \etal 2004a,b), so
that they are likely not Type II QSOs similar to CXO~52.

They, however, may be low equivalent width Type II QSOs. Particularly
interesting is GLARE\#3001, which lies directly on the predicted locus
for the low equivalent Type II QSO in Figure~\ref{fig:iz-zJ_idrop}. Assuming
the emission line is \Lya at $z=5.79$ (Stanway \etal 2004a),
$(i_{775}-z_{850})_{\rm AB}=1.62$ is predicted, while $1.66\pm0.20$ is
measured. Likewise, $(J-K_s)_{\rm AB}=1.13$ is predicted while
$0.89\pm0.45$ is measured (Eyles \etal). It should be noted that the
absence of a detectable \NV$\lambda1240$ line does not preclude a Type
II QSO, as the \Lya to \NV flux ratio can be very large (15--30) in
Type II QSOs (Stern \etal 2002; Mainieri 2005). The possible presence
of \Hb and \OIII$\lambda\lambda4959, 5008$ would also contribute to the
measured Spitzer flux at 3.6$\mu {\rm m}$. This object thus appears a
prime candidate for a low equivalent width Type II QSO.

The presence of the $(i_{775}-z_{850})_{\rm AB}>1.3$ spurs raises the
question of whether some moderate redshift high equivalent width Type
II QSOs could masquerade as high redshift starbursts in future
surveys. The \CIV$\lambda1549$ line at $z=4.5$ could be mistaken for
\Lya at $z=6.0$. Searches for a Balmer break between the $K_s$ band
and $3.6\mu {\rm m}$, as was done using {\it Spitzer} data by Eyles
\etal, could reveal a false break due to the presence of \Ha at
$\lambda=3.6\mu {\rm m}$. The models of Kwan \& Krolik predict \Ha
should be $2-6$ times as strong \Hb, which is a strong line in CXO~52,
with about 10\% the flux of \Lya (Stern \etal).

\begin{figure}
\begin{center}
\leavevmode \epsfxsize=3.3in \epsfbox{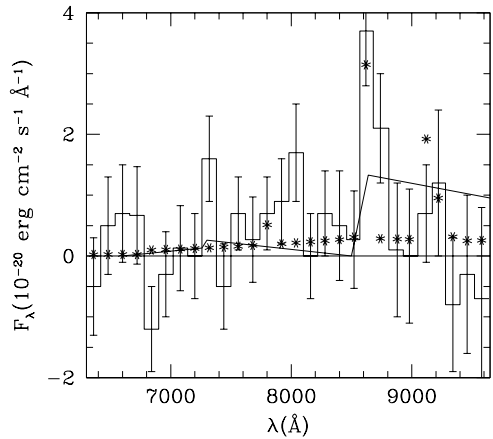}
\end{center}
\caption{A low resolution ($R=100$) grism spectrum (histogram) of
object 3317 in the GRAPES survey from Malhotra \etal (2005), with
their best-fitting model galaxy spectrum (solid line). The points show
the spectrum of CXO~52, redshifted to $z=4.6$ and smoothed and
rebinned to the same resolution and sampling as the grism
spectrum. The emission line at $\lambda\approx8600$\AA\ is due to
\CIV~$\lambda1549$. It is assumed the \Lya emission line has been
absorbed (see text).
}
\label{fig:QII-obj3317}
\end{figure}

Recently, Malhotra \etal (2005) identified a possible large
overdensity of galaxies at $z=5.9$ based on slitless low resolution
($R=100$) grism spectra of $i$-drops in the UDF as part of the GRAPES
survey (Pirzkal \etal 2004). The spectra of the 29 sources with
$(i_{775}-z_{850})>0.9$ they identify fail to show clear evidence for
the presence of any Type II QSOs with strong \Lya emission. One of the
sources, object 3317, does however appear to show at least one strong
emission line. The source has $(i_{775}-z_{850})\approx1.1-1.2$ and
lies well away from the predicted colour curves for galaxies (their
Figure 2). The grism spectrum for the source (adapted from their
Figure 6) is shown in Figure~\ref{fig:QII-obj3317} along with their
best-fitting model galaxy spectrum. The spectrum of CXO~52, redshifted
to $z=4.6$, is also shown, smoothed to $R=100$ and rebinned to match
the grism spectrum. The emission line at $\lambda\approx8600$\AA\ is
due to \CIV~$\lambda1549$. The spectrum was rescaled to give the
minimum $\chi^2$ match to the spectrum, and the effect of
intergalactic attenuation was included. A value of $\chi^2=26$ for 27
degrees-of-freedom is found, an improvement over $\chi^2=35$ found for
the model galaxy fit. The \Lya emission line has been removed,
assuming it has been suppressed, either by dust absorption internal to
the galaxy or possibly by damped absorption by a nearby intervening
damped \Lya system. (Such $z_{\rm abs}\approx z_{\rm em}$ damped
systems are known in absorption line surveys, suggesting the QSO may
be part of a galaxy group; eg, Sargent, Boksenberg \& Steidel
1988. There is evidence that some low redshift Type II QSOs do reside
in groups or clusters; eg, Iwasawa \etal 2005.) The object thus
appears a candidate for a Type II QSO.

\subsection{Predictions for the {\it James Webb Space Telescope}}

\begin{figure}
\begin{center}
\leavevmode \epsfxsize=3.3in \epsfbox{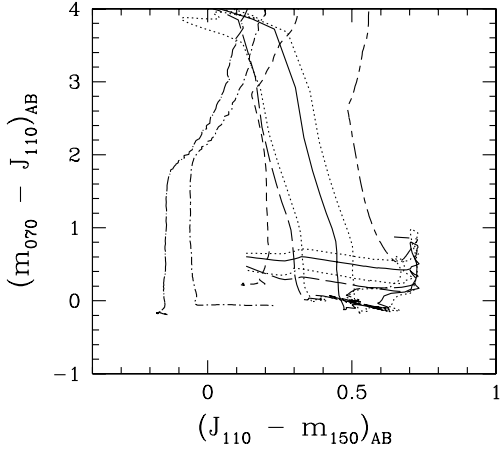}
\end{center}
\caption{$(m_{070}-J_{110})_{\rm AB}$ vs $(J_{110}-m_{150})_{\rm AB}$
colour-colour diagram 
predicted tracks at $z>3$ for starburst
galaxies of age 30~Myr (dotted long-dashed line) and 600~Myr (dotted
short-dashed line), Type I QSO (short-dashed line), and Type II QSO
with high equivalent width emission lines (solid line). The dotted
lines show the tracks for the Type II QSO with the metal lines boosted
or reduced by 50\%. The long-dashed line shows the track for the Type
II QSO with \CIV enhanced by a factor of 2. The short-dashed
long-dashed line shows the predictions for a low equivalent width Type
II QSO. The tracks all lie near each other. Reddening
of the starburst spectra will further compress the locus of tracks.
}
\label{fig:m7J-Jm15}
\end{figure}

\begin{figure}
\begin{center}
\leavevmode \epsfxsize=3.3in \epsfbox{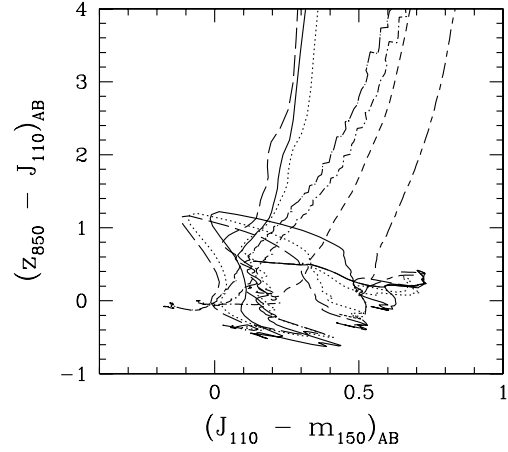}
\end{center}
\caption{$(z_{850}-J_{110})_{\rm AB}$ vs $(J_{110}-m_{150})_{\rm AB}$
colour-colour diagram
predicted tracks at $z>3$ for starburst
galaxies of age 30~Myr (dotted long-dashed line) and 600~Myr (dotted
short-dashed line), Type I QSO (short-dashed line), and Type II QSO
with high equivalent width emission lines (solid line). The dotted
lines show the tracks for the Type II QSO with the metal lines boosted
or reduced by 50\%. The long-dashed line shows the track for the Type
II QSO with \CIV enhanced by a factor of 2. The short-dashed
long-dashed line shows the predictions for a low equivalent width Type
II QSO. The high equivalent width Type II QSO track
with boosted metals lies blueward of the non-boosted track in
$(J_{110}-m_{150})_{\rm AB}$ for $(z_{850}-J_{110})_{\rm AB}<2$ and
redward for $(z_{850}-J_{110})_{\rm AB}>2$. The high equivalent width
Type II QSO tracks are well separated from the Type I QSO, low
equivalent width Type II QSO and starburst tracks.
}
\label{fig:zJ-Jm15}
\end{figure}

The faint magnitude limits that will be achievable by the {\it James
Webb Space Telescope} ({\it JWST}) should make possible the detection
of substantial numbers of high redshift Type II QSOs, if they are
present.

In this section, predictions for the colours detectable by the {\it
JWST} are made. Because the template spectrum does not extend much
into the restframe infra-red, the bands considered are restricted to
$F070W$, $F110W$ and $F150W$. In Figure~\ref{fig:m7J-Jm15}, the
$(m_{070}-J_{110})_{\rm AB}$ vs $(J_{110}-m_{150})_{\rm AB}$ colours
for the Type I and II QSO and starburst spectra are shown. The tracks
are not well-separated. Reddening of the starbursts will further
compress the tracks in colour-colour space. A much better
discriminator is $(z_{850}-J_{110})_{\rm AB}$ vs
$(J_{110}-m_{150})_{\rm AB}$, shown if Figure~\ref{fig:zJ-Jm15}. The
Type II QSO tracks based on CXO~52 are now well separated from the
Type I QSO and starburst tracks. Since the starburst tracks lie
redward of the Type II QSO tracks, redenning of the starbursts will
not lead to confusion with such Type II QSOs. The predictions for the
Type II QSOs are also not very sensitive to the assumed metallicity or
\CIV to \CIII] emission ratio, so that such Type II QSOs should be
clearly identifiable once they become $z_{850}$-dropouts. The low
equivalent width Type II QSO, however, lies near the Type I QSO track
and redward of the starburst tracks, so such Type II QSOs would still
be difficult to identify uniquely based only on their colours.

\subsection{Predictions for UKIDSS}

\begin{figure}
\begin{center}
\leavevmode \epsfxsize=3.3in \epsfbox{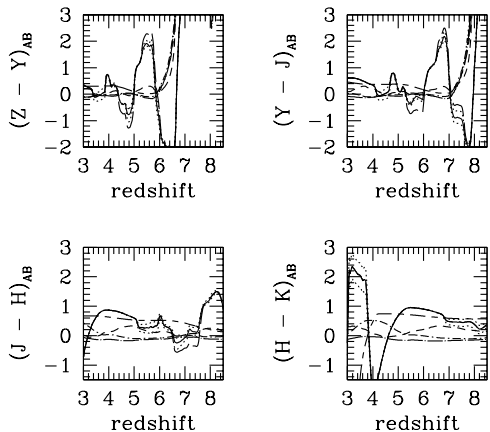}
\end{center}
\caption{Colour evolution of sources in the UKIDSS {\it ZYJHK} filter
system predicted at $z>3$ for starburst galaxies of age 30~Myr (dotted
long-dashed line) and 600~Myr (dotted short-dashed line), Type I QSO
(short-dashed line), and Type II QSO with high equivalent width
emission lines (solid line). The dotted lines show the tracks for the
Type II QSO with the metal lines boosted or reduced by 50\%. The
long-dashed line shows the track for the Type II QSO with \CIV
enhanced by a factor of 2. The short-dashed long-dashed line shows the
predictions for a low equivalent width Type II QSO.
}
\label{fig:UKIDSS_zev}
\end{figure}

The UKIRT Infrared Deep Sky Survey (UKIDSS, Lawrence \etal 2005, in
preparation) covers the wavelength range $0.83-2.37 \mu$m, including a
new $Y$-band covering $0.97-1.07 \mu$m, and is expected to continue
over the next seven years. The Ultra Deep Survey will reach to (AB)
$K\approx25$ and $J\approx26$ (S. Warren private communication). This
is deep enough to detect a source like CXO~52 out to $z=5-6$
(Figure~\ref{fig:magev}).

The evolution of the colour combinations $Z-Y$, $Y-J$, $J-H$ and $H-K$
is shown in Figure~\ref{fig:UKIDSS_zev} for the various sources
considered here. Of particular use for identifying Type II QSOs with
large equivalent width emission lines are $Z-Y$ and $Y-J$. Type II
QSOs in the redshift range $4.5<z<5$ will appear unusually blue in
$Z-Y$, those in the redshift range $5<z<6$ will appear unusually red
in $Z-Y$, and those in the redshift range $6<z<7$ will appear
unusually blue in $Z-Y$ and red in $Y-J$. Those at $7<z<8$ will appear
as $Z$-dropouts that are unusually blue in $Y-J$, of which those with
$z>7.5$ will appear unusually red in $J-H$. The $H-K$ colours also
suggest lower redshift Type II QSOs may be selected as unusually red
for $3<z<4$ and unusually blue for $4<z<4.5$; however, as noted above,
the $K$-band magnitude predictions are very uncertain for $z\lsim5$.

\section{Summary and discussion}

Spectra for Type II QSOs are synthesized based on observed spectra and
photometric measurements, and colours predicted at higher
redshifts. The possibility that some Type II QSOs have high equivalent
widths suggests unusual colours may be expected for these objects,
distinct from Type I QSOs and starbursts. It is shown that the colours
of the $z_{850}$-dropouts discovered in the UDF by Bouwens \etal are
matched by Type II QSOs, as are some of the $i_{775}$-drops found in
the UDF and GOODS. In the case of $i_{775}$-drops, the large but
measurable $(i_{775}-z_{850})_{\rm AB}$ colour is achievable in only
narrow redshift windows, one at $z\gsim6$ when the objects abruptly
become $i_{775}$-dropouts, and a second near $z=4.6$ for which
$(i_{775}-z_{850})_{\rm AB}>1.3$.

It may seem unlikely to find such objects given their rarity at lower
redshifts. It should, however, be noted that even at low redshifts,
Type II AGN appear to be a factor of a few to as much as an order of
magnitude more abundant than Type I AGNs at low luminosities
(Mart\'inez-Sansigre \etal 2005; Simpson 2005). The luminosity
function of low-luminosity Type I QSOs at high redshifts is
unknown. An estimate is made by Meiksin (2005a), under the assumptions
of either pure luminosity evolution (PLE) or pure density evolution
(PDE), based on the low luminosity QSO counts of Hunt \etal (2004) at
$z\approx3$ and the bright QSOs detected at high redshift ($3.6\lsim
z\lsim6$) by the SDSS (Fan \etal 2001, 2004). Using the maximum
likelihood PLE model of Meiksin (2005a) (as given in his Table~1), the
predicted number of Type I QSOs at $z>5.7$ in the $11.5\,{\rm
arcmin}^2$ of the UDF with $z^\prime<28$ is about 0.07 (0.002 in the
PDE model, as the UDF should probe well below the knee in the
luminosity function in this model). Thus the number of detectable Type
II QSOs in the UDF is expected to lie in the range $0.01-1$. Within
the $165\,{\rm arcmin}^2$ of the GOODS-S ACS field, the number of
$z^\prime<27$ Type I QSOs at $z>5.7$ is predicted in the PLE model to
be 0.6 (0.03 in the PDE model), and a number per unit redshift of
$dN/dz\approx1$ for $z\approx4.7$ (0.1 in the PDE model). Thus
detecting at least one Type II QSO is possible, although several may
be unlikely. But this is one reason why searching for them is of such
interest. For instance, if most galaxies went through an AGN phase
early in their histories, then, in AGN unification scenarios, most
would appear as dim Type II objects. If they went through the AGN
phase at $z\gsim6$, then their numbers may be larger than the above
estimates, and a few to several may be detectable in the UDF and GOODS
fields, sufficient to provide a subtantial fraction of the photons
required to re-ionize the IGM (Meiksin 2005a).

Although the UKIDSS Ultra Deep Survey is a somewhat shallower survey,
the relatively large survey area of $0.78\,{\rm deg}^2$ yields a
detectable number of Type I QSOs:\ about six are predicted at $z>5.7$
for $J<26$ by the PLE model, and $dN/dz\approx13$ at
$z\approx4.7$. Several Type II QSOs may thus be detectable at these
redshifts. While the predicted numbers of Type I QSOs are
substantially reduced under the PDE model, about 0.4 at $z>5.7$ and
$dN/dz\approx1.6$ at $z\approx4.7$, the Ultra Deep Survey should prove
an effective means for determining which of these models more closely
describes the actual evolution of QSOs.

Without follow-up spectroscopy, it is difficult to distinguish a
starburst from an AGN. The images of the band-dropout objects tend to
be compact (eg, Stanway \etal 2004a), as would be expected for \HII
regions, but also for an AGN, so images may not readily be used to
distinguish starbursts from AGNs. Extended emission would not necessary
preclude an AGN nature either, both because the emitting regions
extend from scales of several to a few tens of kiloparsecs in size
(Hines \etal 1999) and because some may be embedded in starbursts and
so have multiple bright emitting regions; both AGN and starbursts are
generally believed to be triggered by merger activity. Even with
spectroscopy, high equivalent width Type II QSOs at $z\approx4.6$
could still masquerade as starbursts, with \CIV emission being
mistaken for \Lya at $z\approx6.0$. Strong \Ha emission, falling at
$\lambda=3.6\mu {\rm m}$, could also mimic the 4000\AA~ break of a
starburst at $z\approx6.0$. Even if \Lya emission is correctly
identified, the absence of \NV emission does not preclude a Type II
QSO, as it would a Type I QSO, since the \Lya to \NV ratio in Type II
QSOs is observed sometimes to be very large.

The absence of x-ray emission also may not necessarily exclude an AGN
nature. The expected x-ray flux appears to cover a broad range for a
given UV flux. The ratio of restframe 5.4~keV to 1915\AA~ fluxes ($\nu
f_\nu$) is 0.45 for CXO~52, and the ratio of restframe 25.7~keV to
1915\AA~ fluxes is 4.6 (Stern \etal). The corresponding values for
CDFS-263 are an order of magnitude lower, with a ratio of restframe
5.8~keV to 1957\AA~ fluxes of 0.05 and of restframe 27.9~keV to
1957\AA~ fluxes of 0.16 (Mainieri \etal). The 2~Ms deep exposure by
{\it Chandra} has a lower flux limit for the detection of sources of
$1.9-9.3\times10^{-17}\,{\rm erg\, cm^{-2}\, s^{-1}}$ in the
$0.5-2$~keV (soft) band and $1.1-7.5\times10^{-16}\,{\rm erg\,
cm^{-2}\, s^{-1}}$ in the $2-8$~keV (hard) band, depending on position
(Alexander private communication as quoted in Stanway \etal 2004b). At
$z\approx6$, this implies a source like CXO~52 would need to have an
AB magnitude at 1.3~$\mu$m (observed) of about $26.5-28$ to be
detectable in the x-ray, while a source like CDFS-263 would need to be
2.5 or more magnitudes brighter. This is comparable to or somewhat
brighter than the $J$- and $H$-band detection limits of the deep
surveys used to detect the band-dropout objects.

Although genuine high redshift band-dropout objects are almost
certainly dominated by starbursts, the possibility that a few are AGN
can have major implications for the re-ionization of the IGM. The
potentially important role played by AGNs in re-ionizing the universe
may be demonstrated by comparing the numbers of UV ionizing photons
generated by black holes and by stars through the lifetime of the
universe. The specific luminosity of an AGN at the Lyman edge is
$L_L\simeq0.1L_{\rm bol}/ \nu_L$, where $L_{\rm bol}$ is the
bolometric luminosity of the AGN and $\nu_L$ is the frequency of the
Lyman edge. The production rate of ionizing photons is then $\dot
N_{\rm bh}\simeq0.2L_{\rm bol}/(h\nu_L)$ (Meiksin 2005a), where $h$ is
the Planck constant. Assuming a conversion efficiency $\epsilon_{\rm
bh}$ of mass into energy for accretion onto a black hole, the global
comoving number density of ionizing photons produced by black hole
accretion in the universe changes at the rate
\begin{equation}
\dot n_{\rm bh} \simeq \frac{0.2\epsilon_{\rm bh}\dot\rho_{\rm bh} c^2}
{h\nu_L},
\label{eq:nbdot}
\end{equation}
where $\dot\rho_{\rm bh}$ is the rate at which the average comoving
mass density of black holes in the universe increases with time
(assumed largely due to accretion). The efficiency of mass conversion
is unknown, but phenomenological estimates are in the range
$0.1\lsim\epsilon_{\rm bh}\lsim0.3$ (eg, Yu \& Tremaine 2002).

The production rate of \HI ionizing photons by stars is sensitive to
the initial mass function (IMF) of the stars and to their
metallicities.  For a Salpeter IMF with metallicity 20\% of solar, the
results of Smith, Norris \& Crowther (2002) correspond to a production
rate of ionizing photons per solar mass of stars formed of
$dN_*/dM\approx10^{61}\,{\rm ph\, M_\odot^{-1}}$. For solar
metallicity, the production rate is a factor of 3 smaller, while it
may be somewhat larger for Pop~III stars. The comoving number density
of ionizing photons produced by stars then changes at the rate
\begin{equation}
\dot n_* \simeq\frac{dN_*}{dM}\dot\rho_*,
\label{eq:nsdot}
\end{equation}
where $\dot\rho_*$ is the rate at which the average comoving mass density of
stars in the universe increases with time.

The number of ionizing photons available to re-ionize the IGM is
reduced by internal absorption, both in an AGN and in a star-forming
galaxy. Denoting the escape fractions from AGN and galaxies by $f_{\rm
bh, esc}$ and $f_{\rm *, esc}$, respectively, the ratio of the
ionization rate of the IGM by black holes to that by stars is
\begin{equation}
\frac{\dot n_{\rm bh}}{\dot n_*}\approx330
\left(\frac{\epsilon_{\rm bh}}{0.2}\right)
\left(\frac{dN_*/dM}{10^{61}\,{\rm ph\,M_\odot^{-1}}}\right)
\left(\frac{f_{\rm bh, esc}}{f_{\rm *, esc}}\right)
\left(\frac{\rho_{\rm bh}}{\rho_*}\right).
\label{eq:ndotratio}
\end{equation}
Here, the approximation is made that $\dot\rho_{\rm
bh}/\dot\rho_*\approx\rho_{\rm bh}/\rho_*$, averaged over the age of
the universe, where $\rho_{\rm bh}$ and $\rho_*$ are the respective
mass densities of QSO black holes and in stars in the universe today.  This
seems a reasonable approximation to make since the inferred growth
time of a central massive black hole in a galaxy is comparable to that
of the stars in the bulge over a broad range of bulge masses, at least
in the present-day universe (Heckman \etal 2004). Estimates for the
current mass densities in QSO black holes and stars are $\rho_{\rm
bh}\approx10^5{\rm M_\odot\,Mpc^{-3}}$ (Yu \& Tremaine 2002)
and $\rho_*\approx3\times10^8{\rm M_\odot\,Mpc^{-3}}$ (Baldry \&
Glazebrook 2003), both for $h=0.7$. The escape fraction of ionizing
photons from AGN is unknown, but their spectra suggest at least half
escape, so $f_{\rm bh, esc}\approx0.5$ is assumed. The observational
upper limit on the escape fraction of ionizing photons from galaxies
is $f_{\rm *, esc}<0.04$ (Fernandez-Soto, Lanzetta \& Chen 2002). This
then gives for the ratio of IGM ionizing photons produced by black
holes to that produced by stars
\begin{equation}
\frac{n_{\rm bh}}{n_*}\gsim
\left(\frac{\epsilon_{\rm bh}}{0.2}\right)
\left(\frac{dN_*/dM}{10^{61}\,{\rm ph\,M_\odot^{-1}}}\right)
\left(\frac{f_{\rm bh, esc}}{0.5}\right)
\left(\frac{f_{\rm *, esc}}{0.04}\right)^{-1}.
\label{eq:nratio}
\end{equation}
Despite the low number of black holes compared with stars, their
higher mass-to-energy conversion efficiency compared with stellar
nuclear fusion and the larger escape fraction from AGN compared with
galaxies make black holes competitive with stars as candidate sources
for the photons which re-ionized the IGM.

The relative ionization rates may be related to the relative fractions
of objects detected in deep surveys as follows. Defining $f_{\rm AGN}$
and $f_{\rm SB}$ to be the fraction of band-dropout objects that are
AGNs and starbursts, respectively, $f_L/f_M$ the intrinsic ratio of
flux densities at the Lyman edge and a fiducial frequency $\nu_M$
normalising the counts of objects, and $\alpha^{\rm eff}_{\rm AGN}$
and $\alpha^{\rm eff}_{\rm SB}$ the respective effective spectral
indices of AGN and starbursts shortward of the Lyman edge, then the
ratio of ionizing photon rates of AGN to starbursts injected into the
IGM is
\begin{equation}
\frac{\dot n_{\rm b}}{\dot n_*}=\frac{f_{\rm AGN}}{f_{\rm SB}}
\frac{f_{\rm AGN, esc}}{f_{\rm SB, esc}}\left(\frac{f_L}{f_M}\right)_{\rm AGN}
\left(\frac{f_L}{f_M}\right)_{\rm SB}^{-1}\frac{\alpha^{\rm eff}_{\rm SB}}
{\alpha^{\rm eff}_{\rm AGN}}.
\label{eq:ndotratio_obs}
\end{equation}
Here, $f_{\rm AGN, esc}$ and $f_{\rm SB, esc}$ are the respective
escape fractions of ionizing photons from the observed AGN and
starbursts. The mean observed AGN escape fraction in particular is
expected to be much smaller than the mean black hole escape fraction
above because the contribution from Type II AGN is expected to be very
small. The only ionizing photons that may be observed from Type II AGN
are those re-emitted from the emission line gas illuminated by the
central engine:\ the direct ionizing photons will be obscured. If only
Type I AGN are counted in Eq.~\ref{eq:ndotratio_obs}, then $f_{\rm
AGN, esc}$ may be taken to refer only to these objects, so that
$f_{\rm AGN, esc}\approx f_{\rm bh, esc}$, although it should be noted
that the amount of re-radiated ionizing photons from Type II AGN is
unknown and may not be completely negligible. It will also be assumed
that $f_{\rm SB, esc}\approx f_{\rm *, esc}$. For Pop~III stars and
hard spectra AGN, $\alpha^{\rm eff}\approx0.5$ is expected, while for
Pop~II stars, $\alpha^{\rm eff}_{\rm SB}\approx1.8-2.3$ is expected
(Meiksin 2005a), while soft spectra AGN may have $\alpha^{\rm
eff}_{\rm AGN}\approx1.8$. Further uncertainty arises from the ratio
of the flux $f_L$ at the Lyman limit to the flux $f_M$ measured most
near the Lyman limit. But these likely combine to an uncertainty of a
factor of only a few. The dominant uncertainty is $f_{\rm bh,
esc}f_{\rm AGN}/ f_{\rm *, esc}f_{\rm SB}$. For escape fractions of
$f_{\rm bh, esc}\approx0.5$ and $f_{\rm *, esc}\approx0.04$, only
about one object in a dozen, with an uncertainty of a factor of at
least a few, need by an AGN for AGN to compete with starbursts as the
dominant source of ionizing photons. A further implication of a
substantial AGN contribution to the re-ionization of the IGM is that
the post-reionization temperature will be substantially boosted if the
AGN are sufficiently hard to ionize \HeII to \HeIII as well, which
current AGN number counts suggest they may be able to do by
$z\lsim5.5$ (Meiksin 2005a).

Ultimately, combining {\it HST} observations of $z_{850}$-dropouts
with future follow-up {\it JWST} imaging to detect high redshift Type
II AGN may be the best means of settling the question of how prevalent
Type II QSOs are at high redshifts.

\section{ACKNOWLEDGEMENTS}

The author kindly thanks Steve Warren for providing response curves
for the UKIDSS filters and for useful comments, as well as Omar
Almaini for helpful comments. The author also thanks the referee
Andrew Bunker for several useful comments and suggestions that
improved this work.


\end{document}